# Energy-Efficient AI over a Virtualized Cloud Fog Network

Barzan A. Yosuf, Sanaa H. Mohamed, Mohamed Alenazi, Taisir E. H. El-Gorashi and Jaafar M. H. Elmirghani
Electronic and Electrical Engineering, University of Leeds, Leeds, United Kingdom

*ABSTRACT*

Deep Neural Networks (DNNs) have served as a catalyst in introducing a plethora of next-generation services in the era of Internet of Things (IoT), thanks to the availability of massive amounts of data collected by the objects on the edge. Currently, DNN models are used to deliver many Artificial Intelligence (AI) services that include image and natural language processing, speech recognition, and robotics. Accordingly, such services utilize various DNN models that make it computationally intensive for deployment on the edge devices alone. Thus, most AI models are offloaded to distant cloud data centers (CDCs), which tend to consolidate large amounts of computing and storage resources into one or more CDCs. Deploying services in the CDC will inevitably lead to excessive latencies and overall increase in power consumption. Instead, fog computing allows for cloud services to be extended to the edge of the network, which allows for data processing to be performed closer to the end-user device. However, different from cloud data centers, fog nodes have limited computational power and are highly distributed in the network. In this paper, using Mixed Integer Linear Programming (MILP), we formulate the placement of DNN inference models, which is abstracted as a network embedding problem in a Cloud Fog Network (CFN) architecture, where power savings are introduced through trade-offs between processing and networking. We study the performance of the CFN architecture by comparing the energy savings when compared to the baseline approach which is the CDC.

## 1 Introduction

In recent times, Deep Neural Networks (DNNs) have revolutionized the foundation for many modern-day Artificial Intelligence (AI) based services in the areas of natural language processing, object detection and robotics, to name a few. DNN models are deployed in myriad of applications from autonomous cars to detection of cancer signs to playing complex games. It is claimed that in many applications, DNN models have surpassed human accuracy [1]. The uptake of AI based applications is ramping up due to the advent of technologies like the Internet of Things (IoT). It is reported that billions of smart objects ranging from RFID tags to smart TVs and vehicles have already been connected to the Internet and their numbers are continually on the rise [2]. IoT alone is expected to produce 5 quintillions of data on a daily basis and driverless cars are reported to generate 4TB of data during a single hour of driving per day [3]. This abundance of sensory data makes AI and DNN models particularly attractive for increased deployments in many areas. However, this huge amount of data is currently transported over the network to large cloud data centers (CDCs) for processing purposes. Processing in CDCs introduces a number of constraints related to QoS metrics such as latency and availability, network bandwidth, privacy and last but not least energy efficiency [4], [5].

Instead, fog computing, as a distributed processing paradigm has been proposed to address the aforementioned constraints by leveraging on the resources located between CDCs and IoT end-devices. The collective resources offered at those intermediate layers are normally neglected in favor of centralized processing using the CDC [6]. Such resources include processing, networking and storage facilities that can be utilized to offload the workload from the CDC. Thus, with fog computing, cloud-based services can be extended closer to the source of data where real-time processing can be performed [7]. Any device that has a CPU and networking capability can act as a fog node [8]. However, despite the advantages fog computing has to offer, there are still many challenging issues that need resolving in order to realize its full potential. These challenges are related to fog networking, QoS, interoperability, computation offloading, resource provisioning and orchestration, to name a few [4].

In this paper, we focus on an energy-efficient resource provisioning approach by virtualizing the resources in a distributed processing network that we refer to as the CFN architecture. This work builds on our earlier proposals in various areas such as distributed processing in the IoT/Fog [9]–[12], green core and DC networks [13]–[22] ,[23]–[28], network virtualization and service embedding in core and IoT networks [29]–[32] and machine learning and network optimization for healthcare systems [33]–[36] and network coding in the core network [37], [38]. Our previous work in [39] dealt with the idea of generic service embedding in an IoT setting, we take the work further in this paper by refining the optimization model and abstracting the virtual service requests (VSRs) that comprise of multiple Virtual Machines (VMs) inter-connected in a virtual topology. This allowed us to scale up the MILP model and represent closer to realistic DNN inference workloads in the optimization framework. In addition, we have considered a more practical network with various overheads such as end-to-end network power consumption for data transport as well as inter-VM communication, Power Usage Effectiveness (PUE) and inter server networking needed to achieve server LANs. These extensions provide interesting trade-offs between processing and networking power consumptions using CFN approach compared to the CDC.

## 2 Related Work

In [40], the authors design a linear optimization model to optimally place virtual IoT services in a distributed network setting, which they refer to as IoT-Cloud network. A number of use cases have been considered to capture different smart environments in the IoT, however, they do not address the impact of the networking power consumption which occurs to establish links between virtual functions i.e., VMs. In another study, the authors of [3], tackle the energy efficiency of machine learning tools by optimizing the software code for devices with limited computation and energy so that it can be deployed on the IoT and edge. In a comprehensive study, the authors of [1] provide a detailed survey and review about the recent advances towards the goal of enabling efficient DNN processing. This involves discussions on the type of hardware and platforms that support DNNs as well as highlighting key trends in reducing the processing cost of DNNs, be it through either hardware changes alone or joint hardware design and DNN algorithm changes.

There has been rising interest in utilizing the collective computation capability of localized devices at the edge, different to the traditional approach where this is neglected in favor of cloud processing. This shift from centralized processing to distributed processing/ fog computing is largely motivated by the increase in the processing capability of contemporary computers and embedded devices such as single-board computers (SBCs). As a result, these devices are capable of processing complex AI and machine learning functions for different applications [41]. The work in [6] studies the distribution of computing over IoT mesh networks. In their work, the authors evaluate energy efficiency through two approaches for realizing a neural network (NN) function on a wireless sensor network (WSN); 1) a centralized approach in which data from sensors that act as neurons must flow to a gateway node for aggregation and processing purposes before an output message is sent to the relevant actuator and 2) network as a computer (NaC) is their proposed approach where neurons are processed on the WSN in a distributed manner. Their results showed that the NaC approach was able to reduce the total energy consumption on sensor nodes as well as improve the network lifetime. Finally, the authors simulate their system model on Cooja and conclude that their proposed approach can be superior to the centralized approach in terms of data loss due to interference.

In another work [42], the authors address the optimal embedding of generic functions motivated by three applications: in network computation, operator placement in distributed databases, and module placement in distributed computing. We believe that our work in this paper is unique and scalable in a number of ways: a) we have considered a practical end-to-end cloud fog architecture in our optimization model in which the input parameters better represent practical scenarios, b) we have designed a scalable optimization framework whereby different dimensions and cost objectives such as function consolidation factor (i.e. constraints on number of VMs processed per node) and latency can easily be added, respectively, and c) in addition to placing inference DNN models, we can easily account for the training model by adding additional related parameters to our optimization framework.

## 2 Virtualized Network Optimization Framework

In this section, we introduce the evaluated network architecture, before explaining the role of the virtualization approach in embedding the VSRs, which represent different DNN models. We conclude this section with the problem formulation, where we present the notations used, key mathematical equations in terms of the objective function and constraints.

### 2.1 Network Architecture

As shown in Figure 1, we consider a Cloud Fog Network (CFN) that comprises of four main layers of processing, namely the IoT end-devices, Access Fog (AF), Metro Fog (MF), and the Cloud Datacenter (CDC). The IoT end-devices are located at the bottom most layer, which is usually called the Edge Network. In the access, we consider a Passive Optical Network (PON), which comprises of several ONU devices aggregating traffic from the IoT layers towards the OLT device. PONs are future proof as they are highly scalable and provide high-bit rates at relatively low costs [43]. We assume that the AF node is connected to the OLT device via low-capacity low end routers and switches. Similarly, the MF node is connected to the aggregation Ethernet switch at the metro. Small organizations or even end-users can deploy their fog units at any point in the IoT-Cloud continuum [44]. The CDC is accessed via the core network. The core network uses IP/WDM technology, and it comprises of two layers, the IP layer and the optical layer. In the IP layer, high core routers are used at each node to aggregate network traffic from the metro network. The Optical layer interconnects the core routers through optical switches and IP/WDM equipment such as EDFAs, transponders, and regenerators [45].

### 2.2 System Model

We assume the CFN architecture depicted in Figure 1 is fully virtualized in that multiple heterogeneous workloads/ VMs can be processed on the considered processing layers regardless of the hardware heterogeneity. We also assume that different DNN algorithms are represented by random virtual topologies that are embedded onto the CFN architecture. As illustrated in Figure 2, a VSR comprising of three VMs is embedded onto the network where only IoT devices are used. The VSR comprises of multiple VMs inter-connected via virtual links. Every VSR has an input VM that must be embedded in the IoT layer only since requests are generated here. Furthermore, if the other VMs are processed on different nodes to the input VM, the networking power consumption incurred due to data transfer can be accounted for. It is important to note that, the considered model of the DNNs represents high-level inference only and not the training as the latter is beyond the scope of this paper. Inference DNN models are pre-trained, hence they are not as computationally intensive as training models because the weights and biases have already been determined [1].

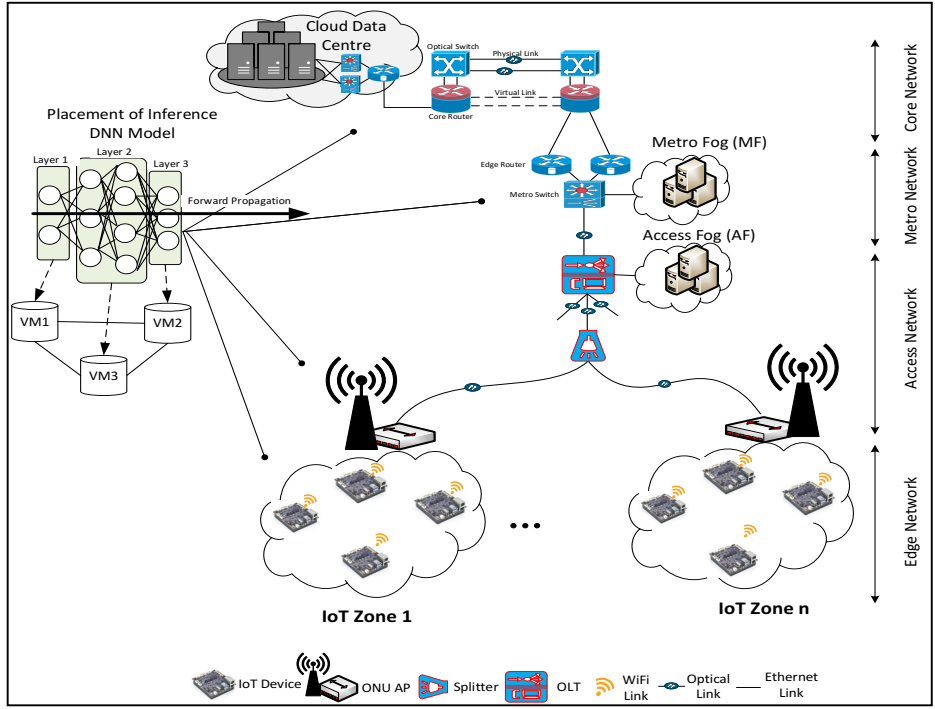

**Figure 1 The proposed Cloud Fog Architecture**

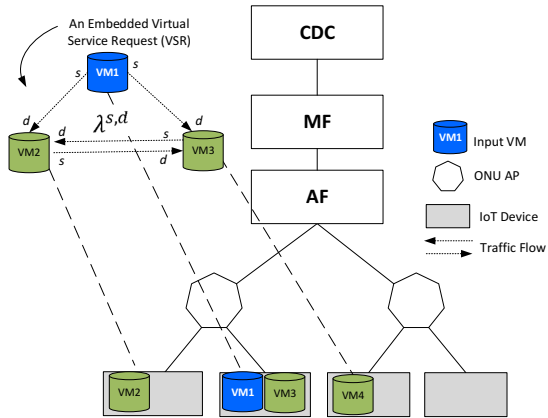

**Figure 2 illustration of VSR embedding across the Cloud Fog Network (CFN) Architecture.**

### 2.2 Problem Formulation and MILP Model

Benefiting from our track record in MILP optimization and particularly in network virtualization and service embedding in [13], [14], respectively, we developed a model to optimize the placement of virtual DNN functions (or VSRs) in the CFN. VSRs comprise of multiple VMs, each VM represents a layer of a DNN model that has a demand for processing (in FLOPS) and networking (in Mbps). Consequently, a VSR is embedded optimally on the CFN model while respecting capacity constraints of processing and networking devices. The physical network shown in Figure 1 is modelled as an undirected graph $G = (N, L)$, where $N$ represents the set of all nodes and $L$ the set of links connecting those nodes in the topology. The VSR $s$ is represented by the directed graph $G^r = (R^r, L^r)$, where $R^r$ is the set of VMs representing virtualized DNN layers and $L^r$ is the set of virtual links connecting those VMs. In Figure 2, we exemplify how demands in a VSR are mapped onto the physical resources in the CFN architecture and show clearly one of the key variables used to establish the virtual links to achieve the inter-VM communication. Before introducing the optimization model, we define the sets, parameters and variables used:

| Sets: | |
|---|---|
| $\mathbb{DC}$ | Set of CDCs. |
| $\mathbb{C}$ | Set of IP/WDM core nodes. |
| $\mathbb{MF}$ | Set of MF nodes. |
| $\mathbb{AF}$ | Set of AF nodes. |
| $\mathbb{O}$ | Set of ONU access points. |
| $\mathbb{I}$ | Set of IoT devices. |

| Symbol | Description |
|---|---|
| $\mathbb{P}$ | Set of nodes that can process a VSR, where $\mathbb{P} = \mathbb{DC} \cup \mathbb{MF} \cup \mathbb{AF} \cup \mathbb{I}$. |
| $\mathbb{R}$ | Set of VSRs. |
| $\mathbb{VM}_r$ | Set of VMs in VSR $r \in \mathbb{R}$. |
| $\mathbb{IP}$ | Set of IoT devices that act as source of demands, where $\mathbb{IP} \subset \mathbb{I}$. |
| $\mathbb{N}$ | Set of all nodes in the CFN architecture. |
| $\mathbb{N}_m$ | Set of neighbor nodes of node m $\in \mathbb{N}$ in the CFN. |
| Parameters: | |
| s and d | Index the source and destination nodes of a VSR topology. |
| b and e | Index source and destination after VSRs are embedded in processing nodes $b, e \in P, b \neq e$. |
| m and n | Index the physical links used to establish the inter-VM communication in the CFN topology. |
| $F^{r,s}$ | Processing requested by node s in VSR r, in FLOPS. |
| $H^{r,s,d}$ | Bitrate requested by VSR r on the virtual link $(s, d) \in \mathbb{VM}_r$. |
| $P_s^r$ | $P_s^r = 1$, if in VSR $r \in \mathbb{R}$, virtual machine $s \in VMr$ is input, otherwise $P_s^r = 0$. |
| $\Pi_n^{(net)}$ | Maximum power consumption of network node n $\in \mathbb{N}$. |
| $\pi_n^{(net)}$ | Idle power consumption of network node n $\in \mathbb{N}$. |
| $C_n^{(net)}$ | Capacity of network node $n \in \mathbb{N}$, in Mbps. |
| $\epsilon_n$ | Energy per bit of network node $n \in \mathbb{N}$, in W/Gb/s, where $\epsilon_n = \frac{\Pi_n^{(net)} - \pi_n^{(net)}}{C_n^{(net)}}$. |
| $\Pi_p^{(LAN)}$ | Maximum power consumption of LAN network inside processing node p$\in \mathbb{P}$. |
| $\pi_p^{(LAN)}$ | Idle power consumption of LAN network inside processing node p$\in \mathbb{P}$. |
| $C_p^{(LAN)}$ | Capacity of LAN network inside processing node p$\in \mathbb{P}$, in Mbps. |
| $EL_p$ | Energy per bit of network node $n \in \mathbb{N}$, in W/Gb/s, where $EL_p = \frac{\Pi_p^{(LAN)} - \pi_p^{(LAN)}}{C_p^{(LAN)}}$. |
| $\Pi_p^{(pr)}$ | Maximum power consumption of a single processing server at node $n \in \mathbb{P}$. |
| $\pi_p^{(pr)}$ | Idle power consumption of a single processing at node $p \in \mathbb{P}$. |
| $C_p^{(pr)}$ | Capacity of single server at processing node $p \in \mathbb{P}$, in GFLOPS. |
| $NS_p$ | Maximum number of servers deployed at processing node $p \in \mathbb{P}$. |
| $E_p$ | Energy per GFLOPS of processing node $p \in \mathbb{P}$. |
| $\delta$ | Proportion of idle power consumed on high-capacity networking equipment. |
| $PUE_n^{(net)}$ | Power Usage Effectiveness (PUE) factor of node $n \in N$ for networking. |
| $PUE_p^{(net)}$ | Power Usage Effectiveness (PUE) factor of node $p \in P$ for processing. |
| Variables: | |
| $\lambda^{b,e}$ | Traffic demand between processing node pair $(b, e) \in \mathbb{P}$ aggregated after all VSRs are embedded. |
| $\lambda_{m,n}^{b,e}$ | Traffic demand between processing node pair $(b, e) \in \mathbb{P}$ aggregated after all VSRs are embedded, traversing physical link $(m, n)$, $m \in \mathbb{N}$ and $n \in \mathbb{N}_m$. |
| $\lambda_n$ | Amount of traffic aggregated by network node $n \in \mathbb{N}$, where $\lambda_n = \sum_{b \in \mathbb{P}} \sum_{e \in \mathbb{P}: b \neq e} \sum_{m \in \mathbb{N}} \sum_{n \in \mathbb{N}_m} \lambda_{m,n}^{b,e} + \sum_{b \in \mathbb{P}} \sum_{e \in \mathbb{P}: b \neq e} \sum_{m \in \mathbb{N}: m \neq e} \sum_{n \in \mathbb{N}_m} \lambda_{n,m}^{b,e}$. |
| $\beta_n$ | $\beta_n = 1$, if network node $n \in \mathbb{N}$ is activated, otherwise $\beta_n = 0$. |
| $\theta_p$ | Amount of traffic aggregated by processing node $p \in \mathbb{P}$. |
| $\Omega_p$ | Amount of workload in FLOPS, allocated to processing node $p \in \mathbb{P}$. |
| $N_p$ | Number of activated processing servers at processing node $p \in \mathbb{P}$. |
| $\Phi_p$ | $\Phi_p = 1$, if processing node $p \in \mathbb{P}$ is activated, otherwise $\Phi_p = 0$. |
| $\delta_b^{r,s}$ | $\delta_b^{r,s} = 1$, if virtual machine $s \in VM_r$ is embedded for processing at node b$\in P$, otherwise $\delta_b^{r,s} = 1$. |

The total power consumption comprises of two parts: A) network power consumption, B) processing power consumption. It is important to note that processing power consumption includes the power consumed by the servers as well as the switches routers within these nodes to provide the LAN.

2.2.1 Network Power Consumption (*net_pc*) which is given by:

$$PUE_n^{(net)} \cdot \left[ \sum_{n \in \mathbb{N}} \epsilon_n \cdot \lambda_n + \sum_{n \in \mathbb{N}} \beta_n \cdot \pi_n^{(net)} \right] \quad (1)$$

The power consumption of the networking equipment comprises of power consumption of routers and switches of all the nodes in the CFN topology depicted in Figure 1 multiplied by the PUE of each networking node.

2.2.2 Processing Power Consumption (*pr_pc*) which is given by:

$$PUE_p^{(pr)} \cdot \left[ \sum_{p \in \mathbb{P}} E_p \cdot \Omega_p + \sum_{p \in \mathbb{P}} N_p \cdot \pi_p^{(pr)} + \sum_{p \in \mathbb{P}} EL_p \cdot \theta_p + \sum_{p \in \mathbb{P}} \Phi_p \cdot \pi_p^{(LAN)} \right] \quad (2)$$

The first term of is the proportional power consumption of the servers whilst the second term calculates the idle power consumption of these servers. The third and fourth terms are the powers consumed by the internal LAN of the processing nodes.

The objective of the MILP is to minimize the total power consumption as follows:

**Minimize:** *net_pc + pr_pc*

**Subject to:**

$$\sum_{b \in \mathbb{P}} \delta_b^{r,s} = 1 \qquad \forall r \in \mathbb{R}, s \in \mathbb{VM}_r: P_s^r \neq 1 \qquad (3)$$

Constraint (3) ensures that all VMs within VSRs are fulfilled, except for input VMs, as these must be mapped to source nodes.

$$\sum_{b \in \mathbb{IP}} \sum_{\substack{s \in \mathbb{VM}_r: \\ P_s^r = 1}} \delta_b^{r,s} = 1 \qquad \forall r \in \mathbb{R} \qquad (4)$$

Constraint (4) ensures that all input VMs are embedded at fixed IoT nodes only.

$$\sum_{n \in \mathbb{N}_m} \lambda_{m,n}^{b,e} - \sum_{n \in \mathbb{N}_m} \lambda_{n,m}^{b,e} = \begin{cases} \lambda^{b,e} & m = s \\ -\lambda^{b,e} & m = d \\ 0 & otherwise \end{cases} \qquad (5)$$

$$\forall b, e \in \mathbb{P}, d \in \mathbb{P}, m \in \mathbb{N}: b \neq e.$$

Constraint (11) preserves the flow of traffic in the network.

Due to space limitations, we have only listed the key constraints in this paper. The remaining constraints deal with binary indicators, capacity constraints on processing and networking devices and establishing physical links between VMs that are connected virtually in a VSR.

## 3  Results and Discussion

In this section, we used the parameters in Table 1 and Table 2 for the processing and networking devices, respectively. It is important to note that, where possible, device parameters have been obtained using equipment datasheets, however, we have also made simple but realistic assumptions. For example, high-capacity networking equipment located in the aggregation point of the access network, metro and core network are used by many applications and services. Hence, we have assumed that, only a portion of the idle power consumption is associated with our application. We assume this to be 3% of the equipment's idle power consumption [9]. We have also assumed that the CDC is a single hop from the aggregation core router (aggregating from metro) with an average distance of 200 km. We assume that in total, there are 20 IoT devices, randomly distributed among four zones; IoT Zone 1 – IoT Zone 4. Each zone is connected via Wi-Fi to an ONU AP and a single OLT aggregates the traffic from these ONUs. As for the workloads, we assume that each VSR has an input VM that must be mapped to the IoT device acting as the source node. In this work, we assume, a single IoT device is acting as the source node. The VM workloads are uniformly distributed between 3-10 GFLOPS and input VMs between 0.1 – 1 GFLOPS. We consider a PUE of 1.25 in AF node, 1.35 in the MF node, 1.12 in the CDC node, 1.5 for core nodes and 1 in remaining nodes [9]. The adopted power profile consists of a proportional and idle part. The proportional part increases with the volume of workload, whilst the idle part is consumed as soon as the device is activated. In the current optimization model, it is assumed that any unused equipment is switched off completely. The MILP model is solved using IBM's commercial solver CPLEX over the University of Leeds high performance computing facilities (ARC3) using 24 cores with 126 GB of RAM [46].

**Table 1 Processing hardware used in the MILP Model.**

| Devices | Max(W) | Idle(W) | GFLOPS | Efficiency (W/GFLOPS) |
|---|---|---|---|---|
| IoT (Rpi 4 B 4GB) | 7.3[47] | 2.56[47] | 13.5[47] | 0.35 |
| AF Server (Intel i5-3427U) | 37.2[47] | 13.8[47] | 34.5[47] | 0.67 |
| MF Server (Intel i5-3427U) | 37.2[47] | 13.8[47] | 34.5[47] | 0.67 |
| CDC (Intel Xeon E5-2640) | 298 [47] | 58.7[47] | 428 [47] | 0.55 |

**Table 2 Networking hardware used in the MILP Model.**

| Devices | Max (W) | Idle (W) | Bitrate (Gbps) | Efficiency (W/Gbps) |
|---|---|---|---|---|
| ONU AP (Wi-Fi) | 15[9] | 9 | 10 | 0.6 |
| OLT | 1940[9] | 60 | 8600 | 0.22 |
| Metro Router Port | 30[9] | 27 | 40 | 0.08 |
| Metro Switch | 470[9] | 423 | 600 | 0.08 |
| IP/WDM Node | 878[9] | 790 | $40/\lambda$ | 0.14 |

As can be seen in Figure 3, four different placement scenarios are evaluated: 1) VSRs being processed at the cloud data center (CDC), 2) VSRs being processed at the access fog (AF), 3) VSRs being processed at the metro fog (MF), and 4) optimizing with MILP the placement of VSRs across the cloud fog architecture (CFN MILP). We observed significant power consumption savings with the CFN approach compared to the CDC, due to local computation in the IoT layer. These savings were 68% on average, 19% in the worst-case scenario and up to 91%. These savings can be attributed to the processing efficiency of the IoT devices as well as the access network used to connect them, hence avoiding various costly overheads such as network power consumption and PUE values associated with the higher processing layers. In the CFN approach, due to the abundance of processing resources, VSRs are always processed by the IoT layer, despite the networking overhead incurred for communication with the source node. What is interesting to note is that the AF and MF nodes are never utilized despite their proximity to the input node at the IoT layer and the negligible network power consumption as per **Error! Reference source not found.**(b) and **Error! Reference source not found.**(c). This is due to the processing inefficiency of these nodes coupled with the high PUE values. In **Error! Reference source not found.**(a), there is a spike in processing and networking power consumption. This is because during very high workloads (20 VSRs), the MILP model chooses to split the workload among the IoT and CDC servers due to capacity violation in the IoT layer. Hence, the CDC node is used to process the excessive demands only and the majority of the total workload is kept at the IoT layer. If the CDC node was to be further away from the aggregating metro node, then the power consumed by the core network may not compensate for the processing efficiency of the CDC, hence processing power consumption will be traded off for networking consumption. In this case both the AF and MF may have a role to play. Another factor that may affect the decision to utilize the intermediary fog nodes instead of the CDC is the type and efficiency of the processing servers deployed as well as how high the PUE value is. We leave these for future research as we believe these changes will introduce interesting insights.

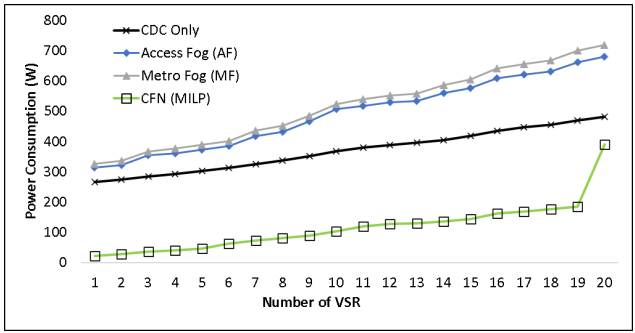

Figure 3 Total Power Consumption

4       Conclusion and *Future Work*

AI and DNN algorithms can enable many applications and services, especially in the IoT area where massive amounts of sensor data must be processed, identified, classified and acted upon. DNN algorithms have surpassed human accuracy in numerous applications, however, this comes at the cost of high computational complexity. In the traditional approach, cloud datacenters alone were the main platform for processing DNNs due to their abundance of processing resources, which can be attractive, however, due to latency drawbacks and energy efficiency, it becomes necessary to evaluate other processing architectures. In this direction, we have studied a CFN architecture where virtualized services representing DNN models can be processed by intermediary servers between the end-devices and the cloud. The results showed that significant savings can be achieved by full utilization of the devices at the IoT layer, whilst fog servers located deeper in the network hierarchy were bypassed in favor of the highly efficient CDC, despite the incurred networking overheads. Motivated by these promising insights, interesting research directions for future work include designing heuristic algorithms of improved computational complexity, considering a realistic core network topology such as the AT&T network, BT network or NSFNET and studying the impact of VM colocation constraints at the IoT layer.

ACKNOWLEDGMENTS
The authors would like to acknowledge funding from the Engineering and Physical Sciences Research Council (EPSRC), INTERNET (EP/H040536/1), STAR (EP/K016873/1) and TOWS (EP/S016570/1) projects. All data are provided in full in the results section of this paper.

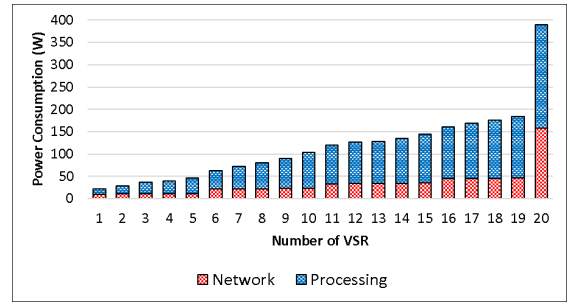

(a)

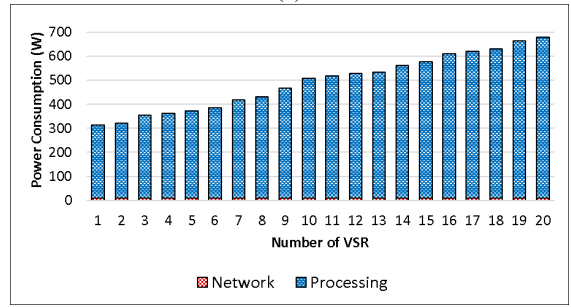

(b)

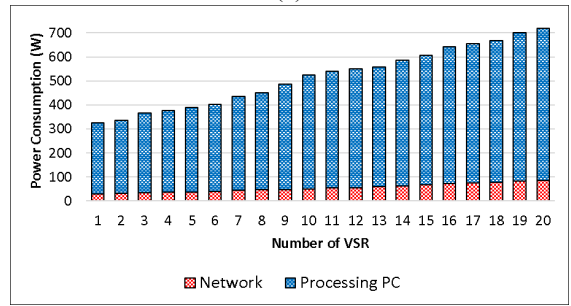

(c)

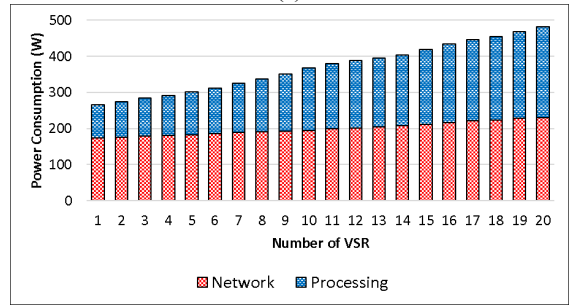

(d)

Figure 4 Network vs. processing power consumption; (a) CFN (MILP), (b) AF and (c) MF (d) CDC.